# Photonic crystal nanocavity based on a topological corner state

Yasutomo Ota[1], Feng Liu[2], Ryota Katsumi[3], Katsuyuki Watanabe[1], Katsunori Wakabayashi[2,4], Yasuhiko Arakawa[1] and Satoshi Iwamoto[1,3]

[1] Institute for Nano Quantum Information Electronics, The University of Tokyo, 4-6-1 Komaba, Meguro-ku, Tokyo, 153-8505 Japan

[2] Department of Nanotechnology for Sustainable Energy, School of Science and Technology, Kwansei Gakuin University, Gakuen 2-1, Sanda 669-1337, Japan

[3] Institute of Industrial Science, The University of Tokyo, 4-6-1 Komaba, Meguro-ku, Tokyo, 153-8505 Japan

[4] National Institute for Materials Science, Namiki 1-1, Tsukuba 305-0044, Japan

Abstract

Topological phonics has emerged as a novel approach to engineer the flow of light and provides unprecedented means for developing diverse photonic elements, including robust optical waveguides immune to structural imperfections. However, the development of nanoscale standing-wave cavities in topological photonics is rather slow, despite its importance when building densely-integrated photonic integrated circuits. In this Letter, we report a photonic crystal nanocavity based on a topological corner state, supported at a 90-degrees-angled rim of a two dimensional photonic crystal. A combination of the bulk-edge and edge-corner correspondences guarantees the presence of the higher-order topological state in a hierarchical manner. We experimentally observed a corner mode that is tightly localized in space while supporting a high Q factor over 2,000, verifying its promise as a nanocavity. These results cast new light on the way to introduce nanocavities in topological photonics platforms.

Faster, denser and more energy-efficient photonic integrated circuits (PICs) have been under intensive development, as witnessed in recent rapid progress in e.g. silicon photonics [1]. A promising approach of the development is to utilize innovative technological platforms, such as those based on photonic crystals (PhCs) [2]. They may enable dense signal wiring as well as excellent control over waveguide group velocity and dispersion [3], although often suffer from non-negligible loss and back reflection, the origins of which are closely linked to unavoidable fabrication imperfections.

In this context, novel PIC platforms based on topological photonics gather great attention due to their potential for robust waveguiding immune to perturbations [4,5]. Indeed, unidirectional, back-reflection-free light propagation has been demonstrated in a microwave quantum Hall system with broken time-reversal symmetry using a gyromagnetic PhC [6]. In addition, helical waveguides have been realized in quantum spin Hall systems based on microring arrays [7], metallic rods [8] and on PhC systems emulating crystalline topological insulators [9,10]. Photonic quantum valley Hall systems have also been investigated and employed for reflection-less waveguiding even under the presence of sharp waveguide bends [11–14]. In contrast to such flourish in topologically-protected waveguides, the development of compact optical resonators in topological photonics is much slower, despite their importance for diverse PIC functions including routing and sorting of signal bits. There are demonstrations of relatively-large topological microresonators, some of which have been employed for topological semiconductor lasers [15–18], but nanoscale stationary photonic cavities still lack in the device list of topological photonics, except for our previous demonstration based on a topological defect in a one dimensional (1D) PhC [19].

For implementing topological photonic nanocavities, the idea of recently-introduced higher order topological insulators (HOTIs) provides a promising starting point [20,21]. The perimeter of a HOTI supports lower dimensional interface states: for the 2D case, topological 0D states are localized at the corners of the HOTI. In this way, HOTIs can naturally introduce stationary cavity modes into the system. HOTIs and topological 0D corner states have been experimentally investigated using microwave circuits [22], phononic crystals [23], crystalline materials [24], electrical circuits [25], acoustic systems [26–28], and optical waveguide arrays [29], but so far not using PhCs.

In this Letter, we report the design and fabrication of a PhC nanocavity based on a topological corner state. We consider a corner of a slab-type 2D PhC that is topologically non-trivial in the sense of its finite two-dimensional Zak phase [30–32], $\boldsymbol{\theta^{Zak}} = \left(\theta_x^{Zak}, \theta_y^{Zak}\right)$. The presence of the corner mode as a higher-order topological state is guaranteed by a combination of the bulk-edge and edge-corner correspondences in a hierarchical manner. The corner state is found to be tightly localized both in space and time and to indeed function as a high $Q$ nanocavity with a small mode volume. We experimentally verify the existence of such nanocavity mode at a corner of the non-trivial PhC. Our results provide an important step for developing topological nanophotonic circuitry that can robustly manipulate light at will in the micro/nanoscale.

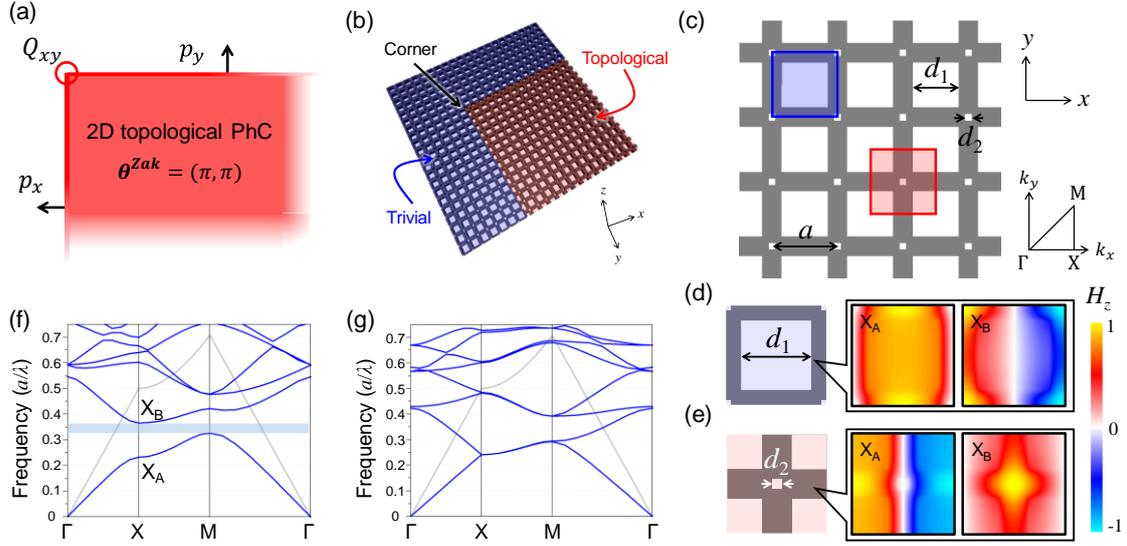

Figure 1. (a) Conceptual schematic illustrating a 2D system supporting both the 1D edge and 0D corner states. (b) Schematic illustration of the investigated 2D PhC. (c) Top view of the PhC, showing how to define the unit cell for the topological (red) and trivial (blue) PhC. Unit cells for the (d) trivial and (e) topological PhC. The insets show magnetic ($H_z$) field profiles simulated at the X point for the lowest and second lowest energy bands. (f) Photonic band structure simulated for the PhC with $d_1 = 0.7a$ and $d_2 = 0.1a$ using the 2D plane wave expansion method. (g) The same in (f) but with $d_1 = d_2 = 0.4a$, showing Dirac points at the X point and quadratic band touchings at the M point.

The celebrated bulk-edge correspondence [33] refers to the connection between topological properties of bulk bands of gapped materials and the existence of edge modes. The simplest case considers the lowest and second lowest energy bands of a gapped 1D material, like the Su-Schrieffer-Heeger model for polyacetylene [34]. The band topology is characterized with a Zak phase [35], defined as the integration of Berry connection within the band over the first Brillouin zone. Given that the unit cell under concern preserves inversion symmetry, a non-trivial Zak phase for the lowest band is associated with the presence of bulk dipole polarization within each unit cell [36]. This in turn results in the emergence of edge polarization, which equal topologically-protected in-gap states localized at the edges of the 1D material. Figure 1(a) schematically describes the situation that we consider here. A 2D PhC with nontrivial Zak phases both in the x and y direction, or with a 2D Zak phase of $\theta^{Zak} = (\pi, \pi)$, is terminated to form a 90-degrees corner. Each component of the 2D Zak phase corresponds to the existence of edge polarization ($p_x$ or $p_y$), or an edge mode at one of the 1D edges [31,37]. In this case, a corner charge ($Q_{xy}$), or a corner mode, is deterministically generated as a convergence of the two sets of the 1D interface polarization, according to the edge-corner correspondence [21,32]. The presence of $Q_{xy}$ is topologically protected in a hierarchy of the bulk-edge and edge-corner correspondence as far as $\theta_x^{Zak} \neq 0 \cap \theta_y^{Zak} \neq 0$. A detailed theoretical description for the emergence of the corner modes is provided in the supplementary material. In what follows, we shall show actual designs of PhCs that translate the situation discussed above.

A schematic of the slab-type topological PhC that we consider here is shown in Fig. 1(b). A topologically-nontrivial 2D PhC (red colored) in a square shape is surrounded by a trivial PhC (blue). Along with the interface between the two PhCs, a 90-degree corner, together with 1D interfaces, is formed. Both the PhCs are based on a common square PhC lattice with a period of *a*, but differ in how to define the unit cell, as shown in Fig. 1(c)-(e). Each unit cell consists of two air holes with different lengths of side ($d_1$ and $d_2$) in a GaAs slab. The two unit cells coincide when shifting the center of one of the unit cells by half a period for both the *x* and *y* directions. As such, the two PhCs support the same band structure in the reciprocal space. Figure 1(d) shows a calculated band structure for $d_1$ = 0.7*a* and $d_2$ = 0.1*a*, using the 2D plane wave expansion method assuming an effective refractive index of the slab to be 2.7. The band structure exhibits a complete 2D photonic bandgap for light with transverse electric polarization. The insets in Figs. 1(b) and (c) show field distributions of the modes of the lowest two energy bands at the X point. For the lowest band of the trivial PhC, an *s*-wave-like, featureless field profile is found, while a *p*-wave-like distribution is supported for its second lowest band. On the other hand, the topological PhC exhibits an apparent band inversion: i.e. a *p*-wave-like mode resides in its lowest band. Since the zero-energy mode at the Γ point always possesses a featureless flat wave function, the *p*-wave-like distribution of the topological PhC suggests the existence of a 'twist' of band nature somewhere within in the momentum space. A band inversion is also observed in the band connecting between the Γ and M points (not shown). These results are reflected in the nontrivial 2D Zak phases of (π, π) for the lowest energy band of the topological PhC. We note that the topological transition observed between the two PhCs occurs when $d_1 = d_2$, at which Dirac points and quadratic band touchings appear respectively at the X and M points, as depicted in the corresponding band structure shown in Fig. 1(e). As far as $d_1 > d_2$, the PhC colored in red remains topological.

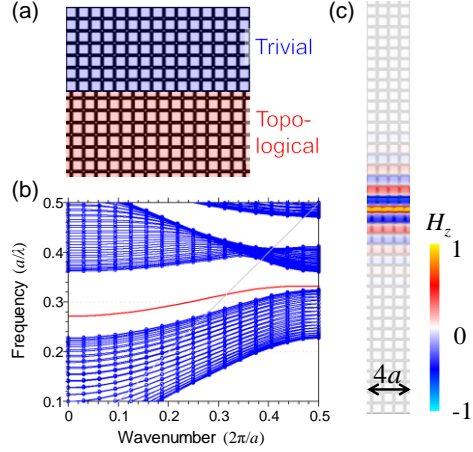

Figure 2. (a) Schematic of the investigated 1D interface. (b) Projected band structure simulated along with the interface. The red and light gray line are of the localized 1D edge mode and of the light line, respectively. The blue curves correspond to those of bulk modes. (c) Field profile for the topological 1D mode localized at the edge, simulated at the Γ point. All the simulation in these plots are performed with the 2D plane wave expansion method.

Before studying properties of the corner, we verify the bulk-edge correspondence in our case by investigating a straight edge of the topological PhC. The situation under the discussion is schematically drawn in Fig. 2(a). Here, the trivial PhC functions as a trivial bandgap material suppressing the leakage of photons from the topological PhC. The abrupt interface between the two PhC forbids the emergence of other modes than that predicted from the bulk-edge correspondence. Along with the interface, we computed a projected band structure as shown in Fig. 2(b). As expected, an in-gap waveguide mode (red line) appears within the mode gap. A part of the dispersion curve lies well below the light line (gray), thus supporting a waveguide mode confined within the slab. Figure 2(c) shows a field profile of the waveguide mode calculated at the Γ point. The localization of the wave at the edge elucidates that the waveguide mode is indeed the edge mode of the topological PhC. Importantly, the 1D topological edge state shows the gapped dispersion curve, which is a prerequisite for the generation of the higher-order topological state [21].

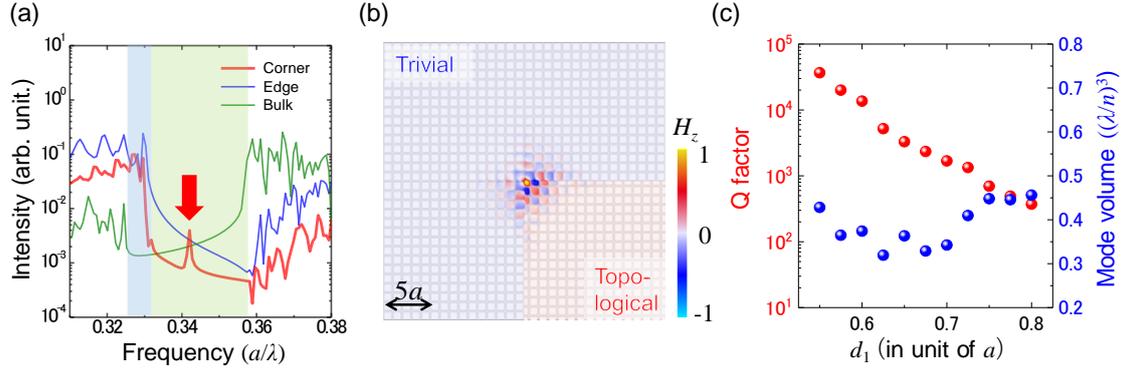

Figure 3. (a) Excitation spectra simulated for the corner (red), edge (blue) and bulk (green) constituted from a PhC with $d_1 = 0.7a$ and $d_2 = 0.1a$. These spectra were calculated using the 2D finite difference time domain method. (b) Magnetic field profile for the mode of the resonance peak at $0.342a/\lambda$, exhibiting the localization at the corner. (c) Evolutions of Q factors (red points) and mode volumes (blue) as a function of $d_1$, while modifying $d_2$ according to $d_2 = 0.8a - d_1$. The slab thickness is assumed to be $0.5a$. The calculations were done with a 3D finite difference time domain simulator.

Now, we numerically investigate light confinement around the corner. Figure 3(a) shows a calculated excitation spectrum around the corner formed with the PhCs of $d_1 = 0.7a$ and $d_2 = 0.1a$. In the simulator, we put a few sources of impulse excitation at various non-symmetric points around the corner and recorded frequency responses of electric and magnetic field components at various locations. The plotted curve is of a typical response. For comparison, we also computed excitation spectra for the bulk PhC and the straight edge and plotted them in the same figure. For the bulk spectrum, we observed a suppression of excitation within the bandgap with a frequency range from $0.326a/\lambda$ to $0.356a/\lambda$, where $\lambda$ expresses a wavelength. A shrunken bandgap is observed for the case of the straight edge, in which the interface waveguide mode exists and partially fulfills the bandgap. For the corner, a sharp resonance peak at a frequency of $0.342a/\lambda$ appears within the remained mode gap. We computed the field profile of the mode causing the peak as shown in Fig. 3(b) and found that it is actually a corner mode as it tightly localizes at the corner. Figure 3(c) shows the evolutions of Q factors and mode volumes when changing $d_1$ and $d_2$ in accordance with $d_1 + d_2 = 0.8a$, computed by the 3D finite difference time domain method. We considered GaAs (refractive index, $n = 3.4$) PhCs with a slab thickness of $0.5a$. For $d_1 = 0.6a$, the corner mode supports a high Q factor of 13,000 and a small mode volume of $0.37(\lambda/n)^3$, demonstrating that it can indeed function as a nanocavity. As far as the photonic bandgap is opened ($0.5a < d_1$), we observed the corner mode with similar mode volumes. Meanwhile, an exponential decrease of Q factor is found as $d_1$ increases. This could probably arise from the weakened confinement in the slab by total internal reflection: the slab becomes more air-like with larger $d_1$ values. The complex behavior of the mode volumes might stem from complicated interplay of the light confinement effects by the photonic bandgap and total internal reflection.

Finally, we experimentally characterize the nanocavity based on the corner mode. We fabricated the designed topological PhC in a 180-nm-thick GaAs slab containing 5 layers of InAs QDs as an internal light probe in the near infrared. The patterning of the structure was done with a combination of electron beam lithography and reactive ion etching. Airbridge PhCs were formed by dissolving a sacrificial AlGaAs layer underneath the GaAs slab. In the current fabrication condition, we found that it is difficult to clearly open small airholes with $d_2 < 0.3a$. Therefore, in the following, we focus ourselves on the study under the vanishing hole size limit of $d_2 = 0$, allowing for avoiding digging excessively-small holes while preserving the topological property of the PhC. Figure 4(a) shows a scanning electron microscope image of a fabricated sample with $d_1 = 0.6a$. A close up image shows the formation of a corner at the interface between the topological and trivial PhC. We then optically characterize the frequency responses of the system by micro-photoluminescence (μPL) spectroscopy at 20 K. Using an objective lens, we focus pump laser light oscillating at 808 nm on the sample. PL signals were collected by the same lens and analyzed by a spectrometer. Figure 4(b) shows observed spectra for the samples designed with $d_1 = 0.6a$. A bunch of peaks for > 1,100 nm is of the Fabry-Pérot resonances originated from the waveguide mode based on the 1D edge states. Indeed, the fringe interval is confirmed to be dependent on the waveguide length (not shown). Separated from the peaks originated from the waveguide modes, we observe a sharp isolated resonance at 1,079 nm, which is likely to stem from the corner mode. The measured frequency separation of the peak from the waveguide mode edge (~ 21 meV) is well comparable with that simulated numerically (24 meV). The linewidth of the peak is deduced to be 0.43 nm by fitting with a Lorentzian function, as shown in Fig. 4(c). The isolated peak supports a relatively-high experimental Q factor of ~2,500. For further confirmation of the origin of the resonance, we investigate position dependence PL intensities as summarized in Fig. 4(c). The resonant mode is tightly localized around the corner both in the $x$ and $y$ directions. The distributions of the mode match well with those calculated, after taking into account the spatial resolution of the μPL setup. When measured another sample of $d_1 = 0.7a$, we again observed a nice agreement between the theory and experiments. From the reasons above, we conclude that we observed the emission from the designed corner modes in our PhCs.

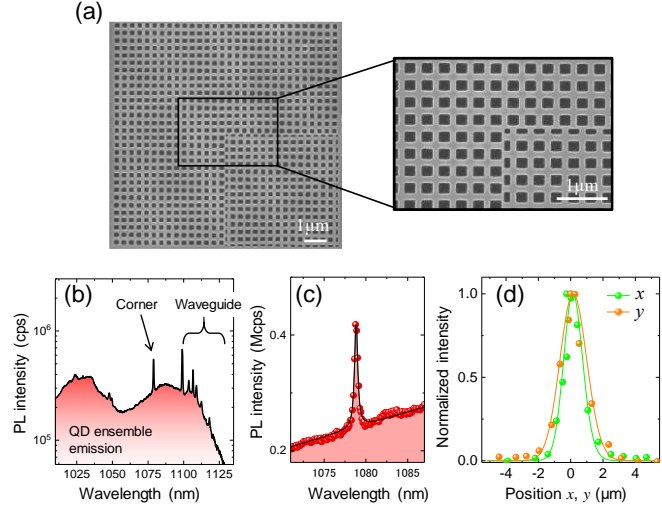

Figure 4. (a) Scanning electron microscope image of a fabricated sample with $d_1 = 0.6a$ and $d_2 = 0a$. The inset shows a close up picture around the corner. (b) Spectrum of the PL emission around the corner. The broad emission peaks widely spanning from 1000 nm to over 1125 nm originate from the QD ensemble. (c) Close-up PL spectrum of the resonance peak at 1078 nm. The black solid line is of fitting using a Lorentzian function with a linearly-increasing background. (d) Summary of position dependent PL measurements around the corner. The green (orange) points show the normalized intensities of the target resonance peak taken along with the $x$ ($y$) direction. The tip of the corner of the topological PhC is taken as the origin of the measurements. The solid lines are of numerical simulations convolved with Gaussian functions expressing the spatial resolution of the PL setup. The spatial resolution for the $x$ direction (~ 1 μm) is better than that of the $y$ direction (~ 1.8 μm), owing to a mechanical slit that is place in front of the spectrometer and transmits signal from a limited range of the $x$ axis.

In summary, we have demonstrated a nanocavity based on a topological corner state. We utilized a slab-type PhC with a non-trivial 2D Zak phase, which is suitable for PIC applications based on transverse-electric polarization. A cascade of the bulk-edge and edge-corner correspondences guarantees the existence of the corner modes, opening a deterministic route to introduce nanocavities into topological nanophotonics platforms. Experimentally, we observed a corner mode with a high Q factor over 2,000, which is tightly localized around the corner of the PhC. These results will be of importance for developing topological nanophotonic PICs with diverse functionalities.

During the completion of the manuscript, we become aware of related works in arXiv [38–41], though none of which measures topological corner states in 2D PhCs in the optical domain.